\documentclass[aps,pra,showpacs,superscriptaddress,twocolumn]{revtex4-2}
\usepackage[utf8]{inputenc}
\usepackage{amsmath}
\usepackage{amssymb}
\usepackage{color}
\usepackage{verbatim}
\usepackage{array}
\usepackage{graphicx}
\usepackage{epsfig}
\usepackage{bm}
\usepackage[ragged]{sidecap}
\usepackage{dsfont}
\usepackage{ulem}
\usepackage{mathrsfs}

\usepackage{caption}
\usepackage{subcaption}
\usepackage{gensymb}
\usepackage{siunitx}
\usepackage{url}
\usepackage{hyperref}
\usepackage{dcolumn}
\usepackage{soul}
\newcolumntype{d}{D{.}{.}{-1}}
\newcolumntype{f}[1]{D{.}{.}{#1}}

\usepackage{hyperref}
\hypersetup{
    colorlinks,
    citecolor=blue,
    filecolor=orange,
    linkcolor=cyan,
    urlcolor=green}

\newcommand{\ie}{{\textit{i.e., }}}
\newcommand{\etal}{\textit{et al.}}

\begin{document}

\title{Absolute measurement of the relativistic magnetic dipole transition in He-like sulfur}

\author{Jorge Machado} 
\email{jfd.machado@fct.unl.pt}
\affiliation{Laboratory of Instrumentation, Biomedical Engineering and Radiation Physics (LIBPhys-UNL),Department of Physics, NOVA School of Science and Technology, NOVA University Lisbon, 2829-516 Caparica, Portugal}

\author{Nancy Paul} 
\email{nancy.paul@lkb.upmc.fr}
\affiliation{Laboratoire Kastler Brossel, Sorbonne Universit\'e, CNRS, ENS-PSL Research University, Coll\`ege de France, Case\ 74;\ 4, place Jussieu, F-75005 Paris, France}

\author{Gabrielle Soum-Sidikov} 
\altaffiliation[current address: ]{IRFU, CEA, Université Paris-Saclay, 91191 Gif-sur-Yvette, France}
%\email{gabrielle.soum@mines-paristech.fr}

\affiliation{Mines ParisTech, Laboratoire Kastler Brossel, Sorbonne Universit\'e, CNRS, ENS-PSL Research University, Coll\`ege de France, Case\ 74;\ 4, place Jussieu, F-75005 Paris, France}

\author{Louis Duval} 
\affiliation{Laboratoire Kastler Brossel, Sorbonne Universit\'e, CNRS, ENS-PSL Research University, Coll\`ege de France, Case\ 74;\ 4, place Jussieu, F-75005 Paris, France}
\affiliation{Institut des NanoSciences de Paris, CNRS, Sorbonne Université, F-75005 Paris, France}

% ** Christophe wants to be removed from the paper
%\author{Christophe Pringent}
%\affiliation{Institut des NanoSciences de Paris, CNRS, Sorbonne Université, F-75005 Paris, France}

\author{Stéphane Macé}
\affiliation{Institut des NanoSciences de Paris, CNRS, Sorbonne Université, F-75005 Paris, France}

\author{Robert Loetzsch}
\affiliation{Helmholtz-Institut Jena, Fröbelstieg 3, 07743 Jena, Germany}
\affiliation{Institut für Optik und Quantenelektronik, Friedrich-Schiller-Universität Jena, Max-Wien-Platz 1, 07743 Jena, Germany}

\author{Martino Trassinelli} 
%\email{martino.trassinelli@insp.jussieu.fr}
\affiliation{Institut des NanoSciences de Paris, CNRS, Sorbonne Université, F-75005 Paris, France}

\author{Paul Indelicato} 
\email{paul.indelicato@lkb.upmc.fr}
\affiliation{Laboratoire Kastler Brossel, Sorbonne Universit\'e, CNRS, ENS-PSL Research University, Coll\`ege de France, Case\ 74;\ 4, place Jussieu, F-75005 Paris, France}
\homepage{http://www.lkb.upmc.fr/metrologysimplesystems/project/paul-indelicato/}

\date{September 30th, 2022}

\begin{abstract}
    We have made the first absolute, reference-free measurement of the $1s 2s \; ^{3}S_{1} \to 1s^{2} \; ^{1}S_{0}$ relativistic magnetic dipole transition in He-like sulfur. The highly-charged S ions were provided by an electron-cyclotron resonance ion source, and the x rays were analysed with a high-precision double crystal spectrometer. A transition energy of \SI{2430.3685+-0.0097}{\electronvolt}  was obtained, and is compared to most advanced bound state quantum electrodynamics calculations, providing an important test of two-electron QED effects and precision atomic structure methods in medium-$Z$ species. Thanks to the extremely narrow natural linewidth of this transition, and to the large dispersion of the spectrometer at this energy, a complementary study was also performed evaluating the impact of different silicon crystal atomic form factor models in the transition energy analysis. We find no significant dependence on the model used to determine the transition energy.
    
\end{abstract}

\maketitle

\section{Introduction}

Tests of bound state quantum electrodynamics (BSQED) are pursued with precision measurements in complementary atomic systems, where the close comparison between experiment and theory allows to disentangle the various facets of quantum electrodynamics. Highly-charged ions (HCI), \ie few electron atomic systems, are a privileged terrain of study as the quantum many body problem can be solved most exactly for these systems, and their strong Coulomb fields lead to amplified BSQED effects in their atomic structure.  Laser spectroscopy of normal and muonic hydrogen allows to test perturbative BSQED to the threshold of third-order effects, and measure the charge radius of the proton \cite{bcrb2022, gmym2020, bvhm2019, fgtb2018}, deuton \cite{pnfa2016} and of the $\alpha$ particle \cite{ksaa2021}, see \cite{kmv2020} for a recent review. High-precision measurements of transition energy, such as the work presented here allow for precision tests of BSQED energy corrections such as self-energy and vacuum polarization or electron-electron and many body relativistic effects, see \cite{ind2019} for a recent review. Precision measurements of other quantities, like Landé $g$-factors in HCI \cite{vdsv2004,swkq2013,wskg2013}, and $g$-factor differences of coupled ions \cite{sdhh2022} also allow to test BSQED contributions, measure fundamental constants \cite{zsks2017} and to place limits on new physics beyond the standard model (BSM).

H-like, single electron systems have been studied across a broad range of species up to Uranium, but far fewer measurements exist for He-like species \cite{ind2019}. A systematic divergence between experiment and BSQED theory for He-like systems has been claimed in the literature \cite{ckgh2012,cpgh2014}, but analysis using the few existing high-precision  results makes it difficult to support such claims \cite{bbkc2007,asgl2012,kmmu2014,esbr2015,bab2015,mssa2018,ind2019}. This is due partly to the fact that, when moving beyond the lightest ions, the transition energies enter the x-ray regime, making direct laser spectroscopy impossible. While new approaches with coherent laser spectroscopy and quantum logic are promising \cite{mlkb2020,ksmw2022}, currently the highest precision method broadly applicable for determining transition energies in HCI is by using crystal spectrometers. These instruments may be coupled with electron beam ion traps (EBIT), electron cyclotron resonance ion sources (ECRIS), and high-energy storage rings depending on the desired atomic number, charge-state, and targeted transition \cite{ind2019}. With these methods, ppm ($\Delta E/E\sim10^{-6}$) accuracy can be achieved for medium-$Z$ species, which allows to probe two-electron QED effects. 

Here we present the first measurement of the $1s 2s \; ^{3}S_{1} \to 1s^{2} \; ^{1}S_{0}$ relativistic magnetic dipole transition energy in He-like sulfur using a double crystal spectrometer without any external reference to theoretical or experimental energy, and compare to the most-advanced BSQED calculations. 

\section{Experimental method}

The measurements were performed at the Laboratoire Kastler Brossel in Paris, using the world-unique experimental setup that couples a high intensity electron-cyclotron resonance ion source (ECRIS) with a double crystal spectrometer (DCS)\cite{gtas2010,assg2014}. The ECRIS, called SIMPA (Source d'Ions Multichargés de Paris) is  jointly operated by the \textit{Laboratoire Kastler Brossel} and the \textit{Institute des Nanosciences de Paris} on the Pierre and Marie Curie campus of Sorbonne University. SIMPA uses \SI{14.5}{\giga\hertz}  microwaves to create an intense plasma of highly charged ions of medium-$Z$ gaseous species, with a source size of a few centimeters, making it well adapted to crystal spectrometers. The  electron temperature inside the ECRIS can reach \SI{46}{\kilo\electronvolt} for light elements like Ar \cite{sags2013}, allowing to create core-excited He-like ions. The trapping depth in the source has been determined based on analysis of Doppler broadening and comparison with simulations to be \SI{\approx 0.2}{\volt} \cite{sags2013}, meaning that the ions have kinetic energies smaller than \SI{\approx 0.2}{\electronvolt}$\times q$, where $q$ is the ion charge. This leads to  Doppler broadening of the emitted transitions of approximately \SI{100}{\milli\electronvolt} \cite{asgl2012}. It should be noted than in precision measurements using EBIT \cite{bbkc2007,kmmu2014}, the depth of the trapping potential is around \SI{200}{\volt}, which lead to a larger broadening. ECRIS do not excite the same lines as EBIT. While the strongest line observed in an EBIT is the $1s 2p\, ^1P_1 \to 1s^2 \,^1S_0$ diagram line, it is the relativistic M1 that is the strongest line observed in ECRIS. The diagram line can also be measured, but is less intense \cite{mssa2018} than the M1.

The DCS uses two \SI{6 x 4}{\centi\meter}, \SI{6}{\milli\meter} thick Si(111) crystals made by the National Institute for Standards and Technology (NIST), whose lattice spacing in vacuum has been measured to a relative uncertainty of \num{0.012 e-6} \cite{assg2014} at a temperature of \SI{22.5}{\celsius} with respect to the definition of the meter. The DCS is thus a reference-free instrument, as the measured wavelengths are directly connected to the definition of the meter. The DCS operates in reflection mode and is capable of attaining world-record precision of a few parts-per-million ($\Delta E/E \sim 10^{-6}$) for few \si{\kilo\electronvolt} x-ray transitions energies emitted by highly-charged ions. The optical axis of the spectrometer is aligned with the source axis so that the DCS can see x rays emanating directly from the plasma. A detailed description of the experimental setup may be found in \cite{asgl2012,assg2014,mssa2018}. 

The experimental campaign was conducted in 2018 and focused on the $1s 2s \; ^{3}S_{1}\to 1s^{2} \; ^{1}S_{0}$ magnetic dipole M1 transition in He-like sulfur. This same transition was measured in Ar during a previous data taking period   \cite{asgl2012}, but as the sulfur transition is located at lower energy, the dispersion in the instrument is higher and here we are thus able to perform a more sensitive measurement. These M1 transitions are unique as they have a natural linewidth that is negligible with respect to the broadening induced by the crystal response of the DCS. Thus, after accounting for the small Gaussian Doppler broadening due to the temperature of the ions trapped in the space charge of the electrons, which has been well characterized \cite{asgl2012,sags2013,assg2014}, the line shape that we obtain for these transitions is directly related to the spectral response of the photon diffracting through the crystal structure. The fundamental description of this phenomenon is provided by dynamical diffraction theory and atomic form factors, the latter which describes the response of the electronic cloud of the Si atoms in the crystal to the incident radiation. 
\begin{figure*}
\centering
    \begin{subfigure}[b]{0.8\textwidth}
        \includegraphics[width=\linewidth]{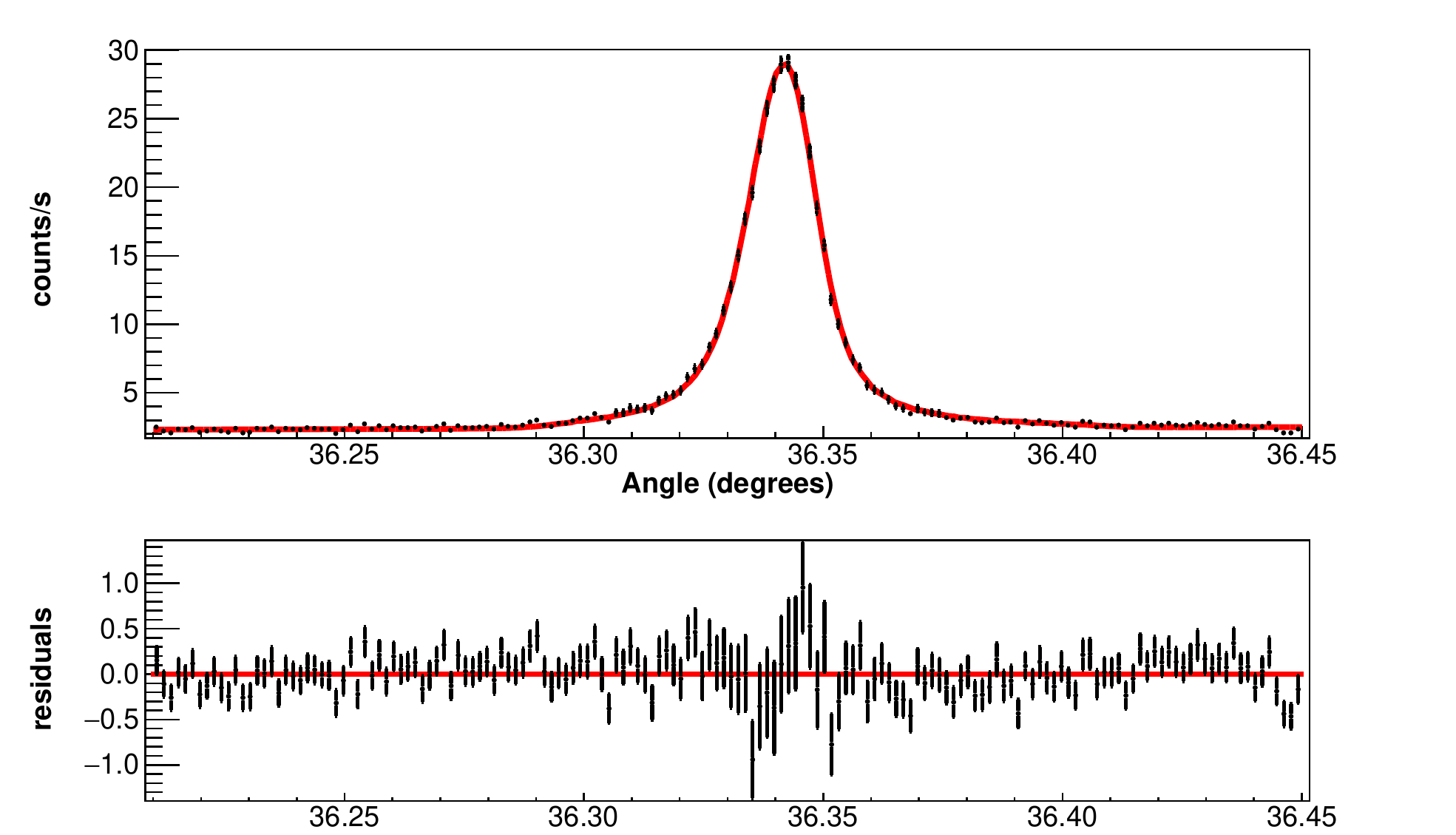}
        \caption{[upper panel] Experimental dispersive spectrum for the $1s 2s \: ^{3}S_{1}\to 1s^{2} \: ^{1}S_{0}$ transition in He-like Sulfur (black points), fit with a simulated response function and linear background (red curve). [lower panel] The fit residuals. }
        \label{fig:S_M1_best_doppler}
    \end{subfigure}
\centering
    \begin{subfigure}[b]{0.8 \textwidth}
       \includegraphics[width=\linewidth]{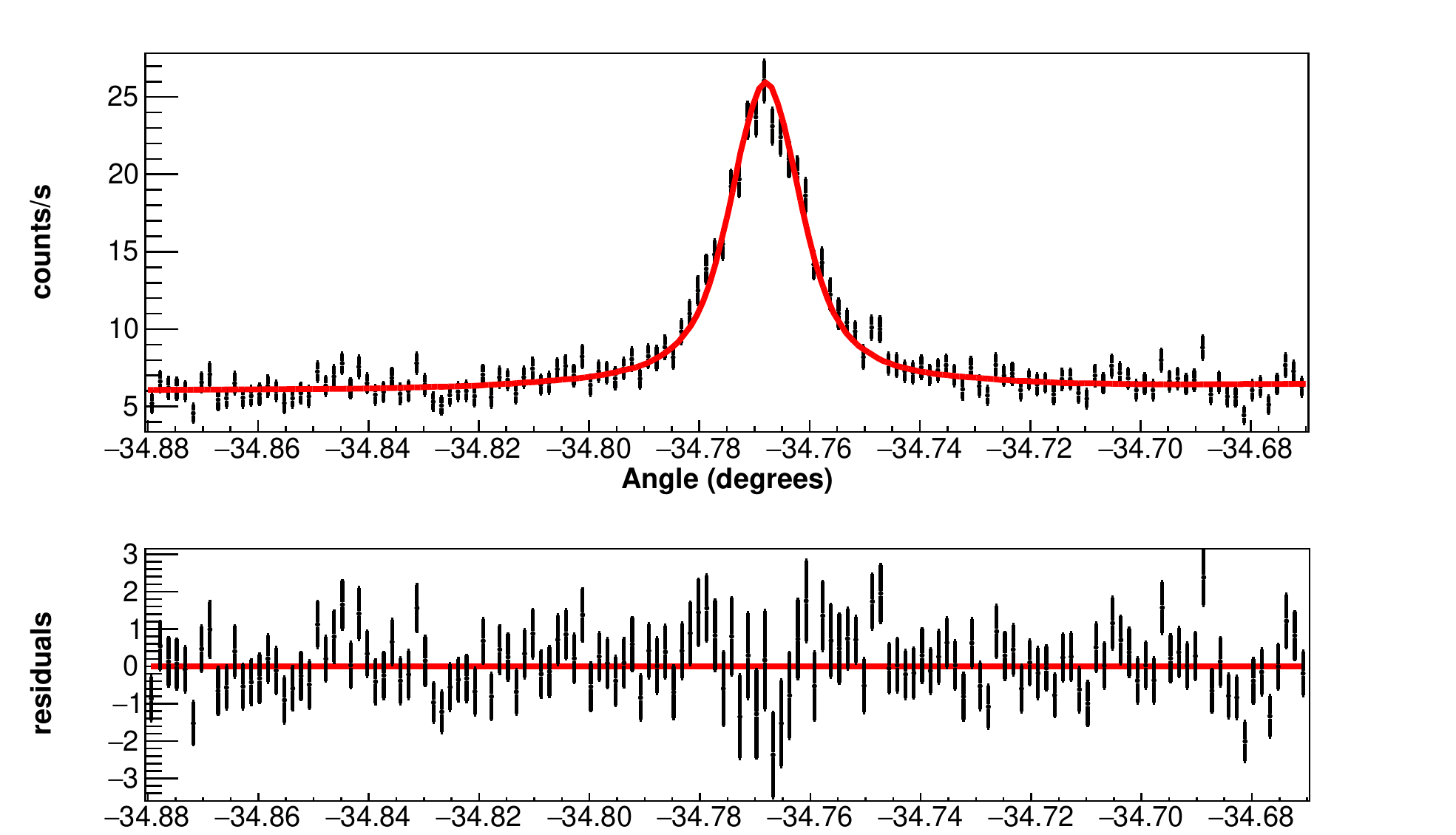}
        \caption{[upper panel] Experimental non-dispersive (parallel) spectrum for the $1s 2s \: ^{3}S_{1}\to 1s^{2} \: ^{1}S_{0}$ transition in He-like Sulfur (black points), fit with a simulated response function and linear background (red curve). [lower panel] The fit residuals. }
        \label{fig:S_M1_parallel}
    \end{subfigure}
\caption{Experimental pair of dispersive  and non-dispersive (parallel) spectra for the $1s 2s \: ^{3}S_{1}\to 1s^{2} \: ^{1}S_{0}$ transition in He-like Sulfur.
\label{fig:ex_spectra_S}
}
\end{figure*}

The measurement and analysis principles have been described in detail elsewhere \cite{assg2014}, and here only the key points will be summarized. During a measurement, the first crystal is maintained at a fixed position, and acts as a first selector in energy as only a small range of photon energies will be reflected onto the second crystal. The second crystal is then oriented in one of two modes: 1) non-dispersive, where the two crystals are parallel or 2) dispersive, where the two crystals deflect x-rays in the same direction.  The peak shape obtained when the two crystals are parallel, \ie in the non-dispersive mode, depends only on the experimental geometry and reflection profile of the crystals. In the dispersive mode the peak shape is a convolution of the instrument response function and of the line shape of the transition. The angular difference between the two modes of the second crystal can be directly connected to the Bragg angle, thus allowing one to analyse the energy of the x rays. A microstepping motor rotate the second crystal continuously within a pre-defined range for each mode, and the number of x~rays is recorded along with the crystal angles and temperatures. During a typical day of measurement, the majority of the time is spent measuring the dispersive-mode spectrum, and non-dispersive-mode spectra are taken at the beginning and end of the day for a given transition. An example of a dispersive and non-disperive mode spectra for the M1 transition in He-like sulfur is shown in Fig.~\ref{fig:ex_spectra_S}.

The analysis of the data is based on an \textit{ab initio} simulation of our spectrometer \cite{assg2014}, which performs exact ray tracing based on the geometry of our setup and of the ECRIS source. The simulation uses as input the reflectivity profiles (commonly called ``rocking curves'') for the Si(111) crystals obtained from an x-ray tracing program based on dynamical diffraction theory. The rocking curves used in the analysis presented here were obtained using the x-ray oriented program (XOP) \cite{sad1998,sad2004,sad2011}. The response functions obtained from this simulation are then used to fit the experimental spectra and determine the transition energies and widths of the measured transitions.

\subsection{Evaluation of the Doppler widths}
\label{sec:doppler}

In the analysis procedure described in detail in \cite{asgl2012, assg2014, mssa2018,mbpt2020}, we must determine first the Doppler broadening of the lines. For the forbidden transition considered here, this is straightforward as the natural width of the line is negligible, so all the broadening seen in the spectra that cannot be explained by the spectrometer response function is due to the Doppler effect. Using the \textit{ab initio} simulation of our setup, first a set of response functions is simulated for the dispersive spectra with different Doppler widths $\Gamma_{\mathrm{G}}^{i}$ ranging from \SIrange{0}{500}{\milli\electronvolt}, assuming the theoretical line energy, setting the temperature in the simulation to \SI[parse-numbers=false]{T= 22.5}{\celsius}, and setting the Lorentzian (natural) linewidth to zero. The dispersive spectra are then fitted with these simulations superimposed on a linear background, using the function 
\begin{equation}
\label{eq:fitfcn}
%symbols that mean something like max or G (for gaussian) must be in roman!!
I\left ( \theta-\theta_0, I_{\mathrm{max}}, a, b \right) =I_{\mathrm{max}} S_{E_{0},\Gamma_{\mathrm{G}}^{i},T_0}(\theta-\theta_{0})+b\theta+a
\end{equation}

\noindent where $S_{E_{0},\Gamma_{\mathrm{G}}^{i},T}$ is the set of simulated response functions with different Gaussian widths, line energy $E_0$ and temperature T. The fitting parameters are the peak intensity $I_{\mathrm{max}}$, the peak centroid $\theta$, and the background slope $b$ and offset $a$. An example fit is shown in Fig.~\ref{fig:S_M1_best_doppler}. 

\begin{figure}[tb]
\includegraphics[width=\linewidth]{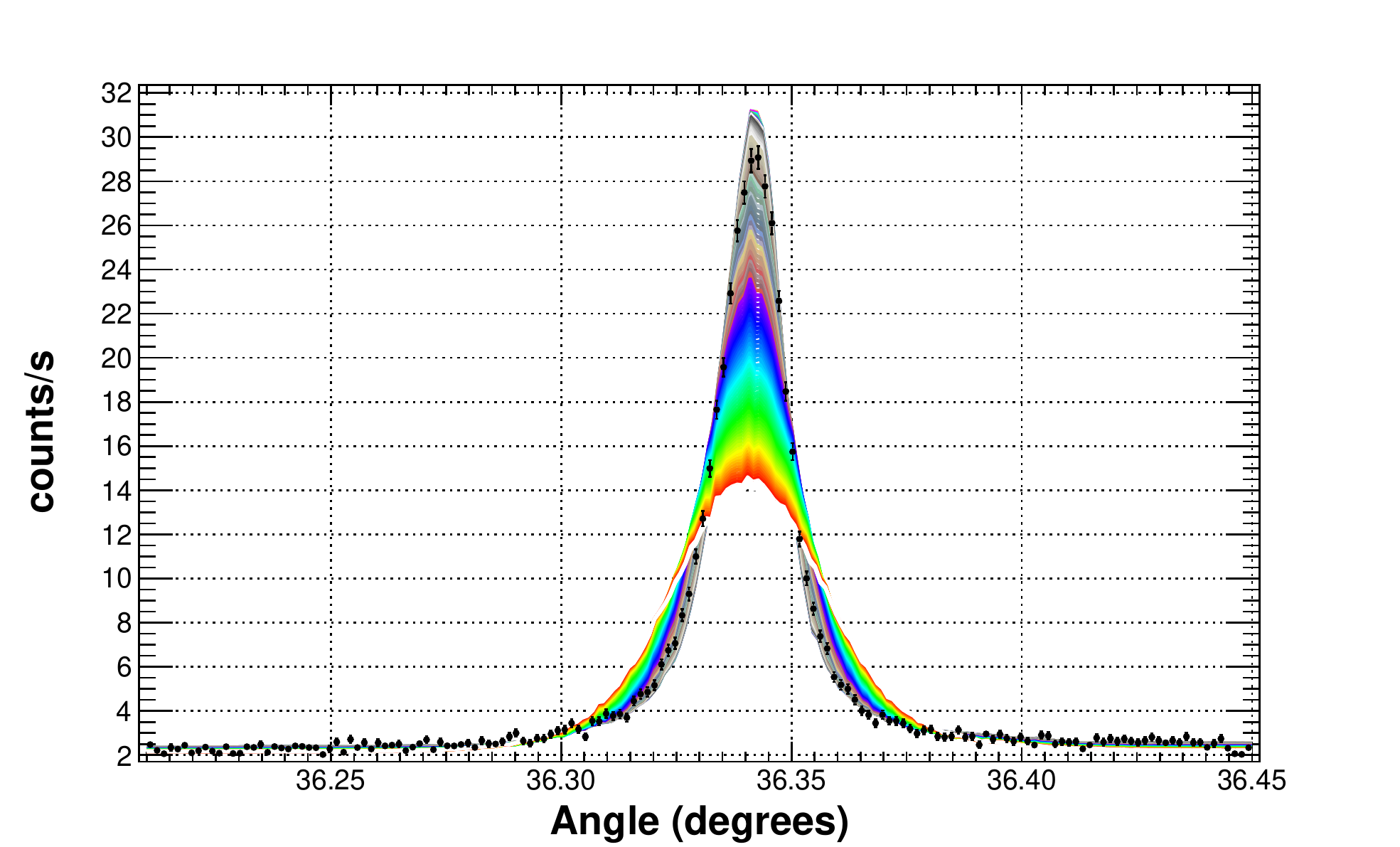}
\caption{Experimental dispersive spectra fit with simulated response functions with different Gaussian Doppler widths, from \SIrange{0}{500}{\milli\electronvolt}. The colors from black to red indicate minimal and maximal simulated Doppler widths, respectively.}
\label{fig:S_M1_nist_doppler_allfits}
\end{figure}

\begin{figure}[tb]
\includegraphics[width=\linewidth]{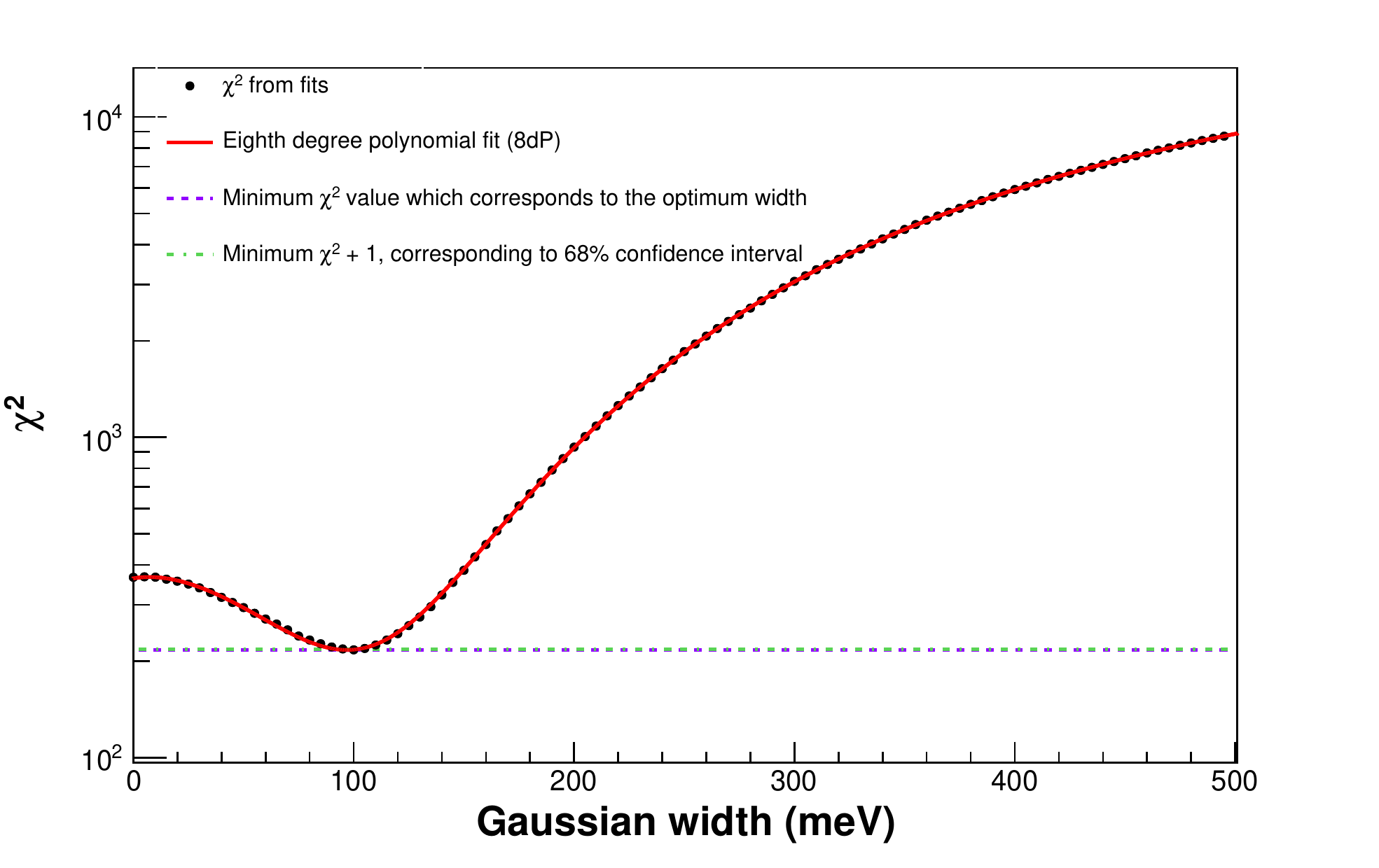}
\caption{Reduced $\chi^{2}$ of the fit to the dispersive spectrum as a function of Gaussian width (black points). The trend has been fitted with an eighth-degree polynomial (red line).}
\label{fig:chi2_doppler}
\end{figure}

\begin{figure}[tb]
%%%% ADD FIGURE HERE
\includegraphics[width=\linewidth]{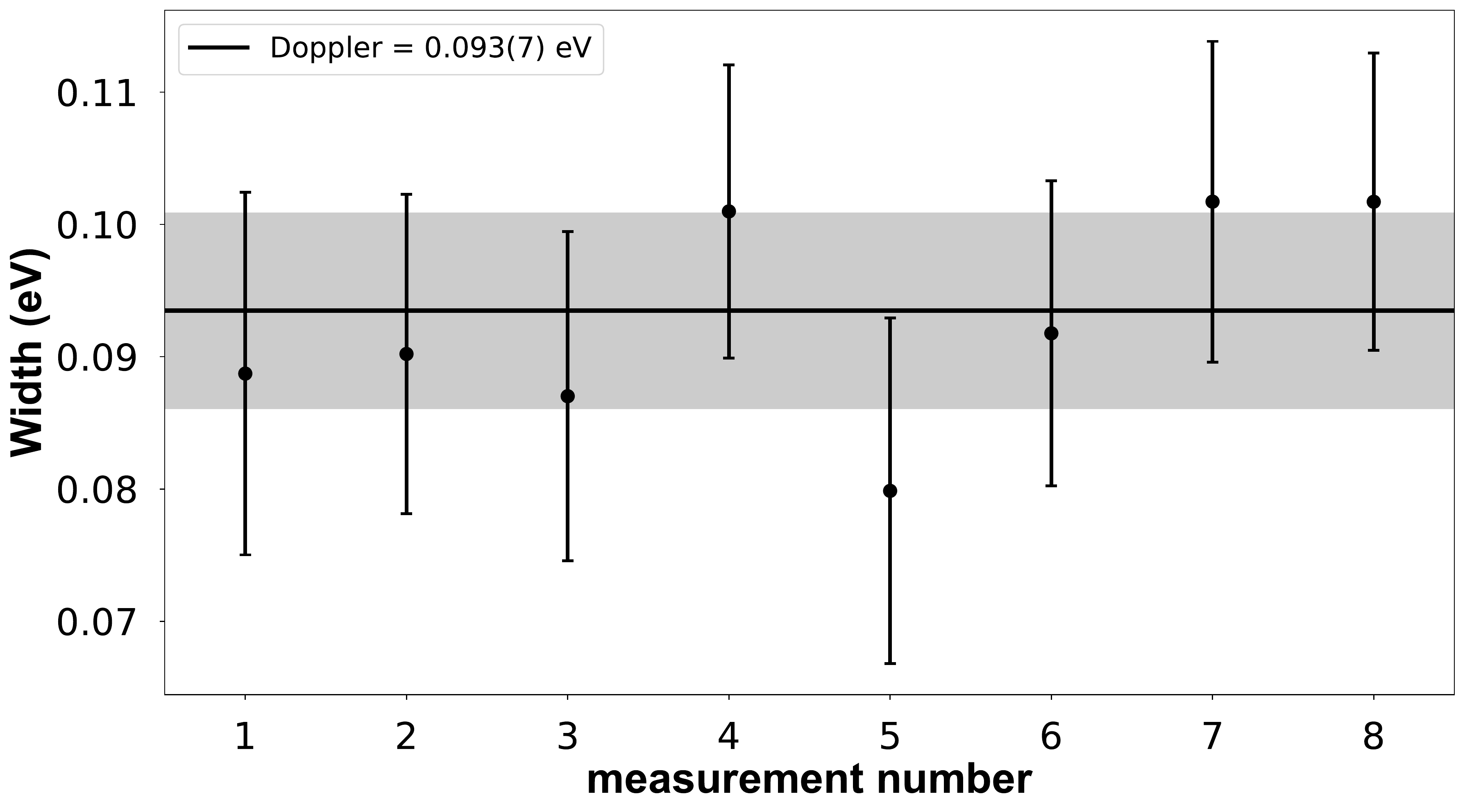}
\caption{Doppler width as extracted from each dispersive spectrum with the statistical error bars. The weighted average with its uncertainty is shown as the solid line and shaded bar, respectively.}
\label{fig:Doppler_avg_xop}
\end{figure}

The quality of the fit is then evaluated by considering the evolution of the reduced $\chi^{2}$ as a function of Gaussian width. Each experimental dispersive spectrum is fitted with the set of simulations with different $\Gamma_{\mathrm{G}}^{i}$, and the minimum in the $\chi^{2}$ curve is determined for each spectrum. Figure~\ref{fig:S_M1_nist_doppler_allfits} shows an example of the fits to a dispersive-mode spectrum with a set of response functions with Gaussian widths in the \SIrange{0}{500}{\milli\electronvolt} range. The $\chi^2$ trend may be seen in Fig.~\ref{fig:chi2_doppler} for the different fits of a single sulfur spectrum. The trend was fitted with an eighth-degree polynomial to obtain the minimum. This analysis was also checked by using the NestedFit \cite{tra2019} Bayesian analysis fitting program, based on the methods described in \cite{tra2017}. In this approach, the log of the Bayesian evidence of each fit to the dispersive spectrum is evaluated as a function of the simulated Doppler width, and the maximum of the evidence generally indicates the maximum likelihood.  The Bayesian evidence, also called marginalized likelihood, is obtained by the integration of the likelihood function over the fit parameters. The (logarithmic) values of the evidence as a function of the Doppler width were then evaluated, and fitted with both an eighth-degree polynomial and splines. The weighted average of the maximum obtained with both polynomial and spline regressions was taken to determine the Doppler broadening from each spectra. The standard uncertainty with this method is given by a $\ln(\mathrm{evidence})$ offset of \num{0.96} from the maximum evidence, which corresponds to the $2\sigma$ confidence range, the accepted standard for Bayesian evidence analysis uncertainty extraction \cite{gat2007}. The evolution of the Doppler broadening obtained in this way for the different Sulfur spectra is shown in Fig.~\ref{fig:Doppler_avg_xop}. The small fluctuations correspond to daily variations in the ECRIS source parameters. 

A weighted average of all the sulfur spectra width was evaluated to obtain an average Doppler broadening of \SI{0.093+-0.007}{\electronvolt}  (FWHM), in agreement with the results of the $\chi^2$ minimization of \SI{0.095+-0.007}{\electronvolt}. This Doppler width was then used for the transition energy analysis described below. Note that this Doppler width is slightly larger than the value published during our analysis of transitions in Li-like sulfur \cite{mbpt2020}, where a value of \SI{0.0917+-0.0074}{\electronvolt} was obtained, though they agree within their statistical uncertainties. We checked the impact of this small change and we found that performing the analysis with the Doppler width fixed at \SI{0.0950}{\electronvolt} instead of \SI{0.0917}{\electronvolt}, leads to changes in the natural linewidths of the $1s2s2p ^{2}P_{1/2,3/2}\to 1s^{2}2s ^{2}S_{1/2}$ transitions of \SI{0.3}{\percent} and \SI{0.8}{\percent}, respectively, negligible with respect to the \SI{17.5}{\percent} and \SI{24.4}{\percent} uncertainties on these linewidths.

\subsection{Evaluation of transition energies}
\label{sec:transE}
\begin{figure}[tb]
%%%% ADD FIGURE HERE
\includegraphics[width=\linewidth]{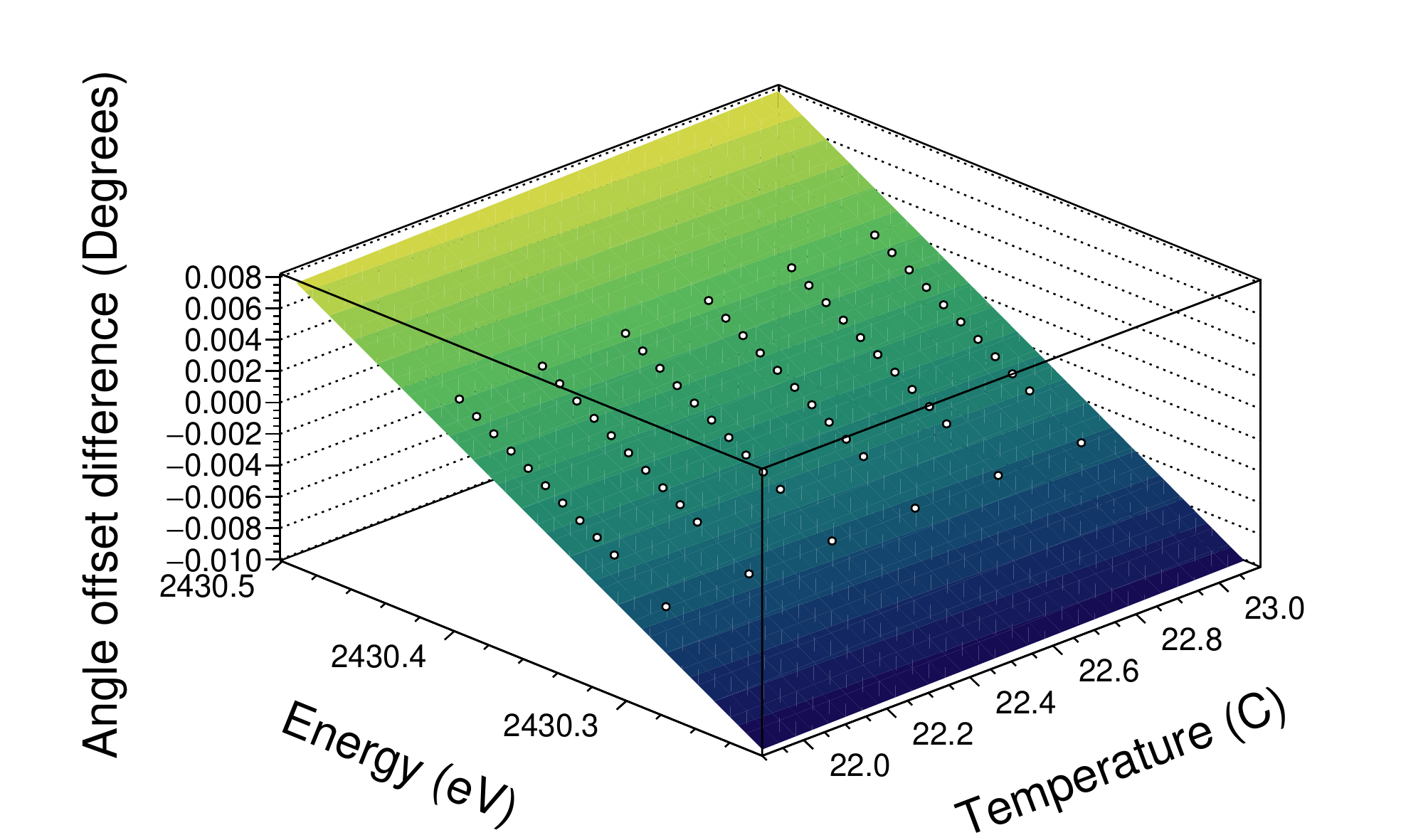}
\caption{Fitted two-dimensional angular offset function from Eq. \eqref{eq:offsets} and experimental results (white spheres), for an example pair of experimental spectra for the sulfur M1 transitions. The fit is performed taking into account the statistical error bars in each point.}
\label{fig:offsetSurfaces}
\end{figure}

With the Doppler broadening determined from the analysis described above, a new set of simulations for both non-dispersive and dispersive modes is performed for different transition energies $E_{i}$ and crystal temperatures $T_{l}$, which both shift the line position. The energies are simulated in a grid around the theoretical transition energy according to $E_{i}=E_{\mathrm{theo}}+i \Delta E$, where $\Delta E$ is an energy step (generally \SI{\approx 10}{\milli\electronvolt}, and $i$ is an integer that can take positive or negative values. Each $E_{i}$ is simulated for different crystal temperatures \SI{20}{\celsius}\SI[parse-numbers=false]{\leq T\leq 25}{\celsius}.  Each experimental spectrum is then fitted according to Eq.~\eqref{eq:fitfcn}, where now $E_0=E_{i}$, $T_{0}=T_{l}$, and the Gaussian broadening $\Gamma_{G}^{i}$ is fixed to the value obtained with the procedure described in Sec.~\ref{sec:doppler}. 

For each experimental pair of dispersive and non-dispersive spectra, corresponding to a day of measurement, the offsets in angle between the simulated and experimental spectra are determined, where the offsets are given by

\begin{equation} 
\label{eq:offsets}
\begin{aligned}
%use \Big( to avoid two different size on the two different parts!
\Delta \theta^{n,k,l}_{\mathrm{Exp}-\mathrm{Sim}}= \Big( \theta^{n}_{\mathrm{Exp}-\mathrm{D}}-\theta^{n}_{\mathrm{Exp}-\mathrm{ND}} \Big)-\\
\Big( \theta^{k,l}_{{\mathrm{Sim}-\mathrm{D}}}-\theta^{k,l}_{\mathrm{Sim}-\mathrm{ND}} \Big).
\end{aligned}
\end{equation}

In the above expression, $n$ indicates a given pair of dispersive ($\mathrm{D}$) and non-dispersive ($\mathrm{ND}$) spectra, resulting from a single day of measurement. If $\Delta \theta^{n,k,l}_{\mathrm{Exp}-\mathrm{Sim}} =0$, the temperature and energy of the simulation correspond to those of the experiment. The angular offset $\Delta \theta^{n,k,l}_{\mathrm{Exp}-\mathrm{Sim}}$ is then evaluated on the grid of simulated energies and temperatures and fitted with the bi-dimensional function:
\begin{equation}
\label{eq:ETfit}
\Delta \theta_{\mathrm{Exp}-\mathrm{Sim}}\left ( E, T \right )= p + qE+rE^{2}+sET+uT+vT^{2}.
\end{equation}

\begin{figure}[t]
\includegraphics[width=\linewidth]{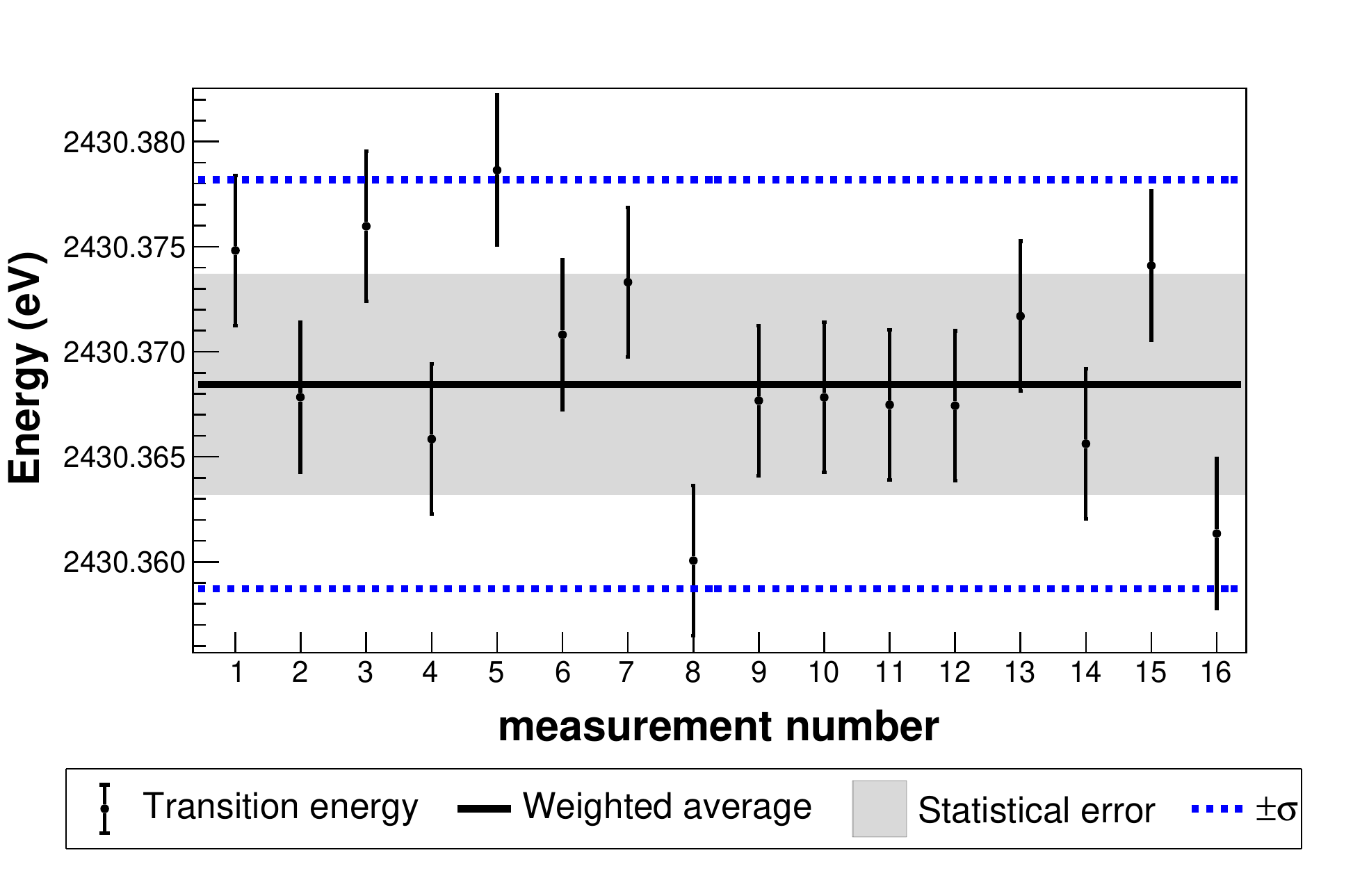}
\caption{ Results of the transition energy analysis for the different pairs of dispersive and non-dispersive spectra recorded during the experiment. The error bars on the points correspond to the statistical, temperature and the angle encoder measure uncertainties added in quadrature. The black solid line represents the weighted average considering only the statistical uncertainty of each point. The gray shaded region corresponds to the weighted standard deviation. The blue dashed lines indicates the standard uncertainty on the weighted average including both statistical and systematic uncertainties.  }
\label{fig:S_M1_En_XOP}
\end{figure}

For each pair of experimental spectra $n$, the above expression is used to obtain the experimental line energy, by determining where $\Delta \theta_{\mathrm{Exp}-\mathrm{Sim}}\left ( E^{n}_{\mathrm{exp}}, T_{\mathrm{exp}} \right )=0$, where $T_{\mathrm{exp}}$ is the measured temperature of the second crystal. An example of an angular offset surface used to obtain the transition energy from a pair of measurement spectra is shown in Fig.~\ref{fig:offsetSurfaces}. The transition energies extracted from each pair of spectra are then averaged to obtain the final transition energy including statistical uncertainties of \SI{2430.3685+-0.0053}{\electronvolt}, as shown in Fig.~\ref{fig:S_M1_En_XOP}.

Table~\ref{tab:syst_uncert} shows the experimental systematic uncertainties and their contributions in \si{\electronvolt} to the uncertainties on the transition energies. Including these effects, the final transition energy is \SI{2430.3685+-0.0097}{\electronvolt}. Note that with respect to our previous works \cite{assg2014,mbpt2020}, the systematic uncertainty associated with the energy to wavelength conversion has been removed, since the new definition of the International System of Units in \num{2018} has defined the Planck constant to be exact \cite{tmnt2021}, as the speed of light has been since \num{1983}. Nevertheless this redefinition has a negligible effect on our systematic uncertainty.
As the energy of the sulfur M1 transition is lower than the one in Ar, the Bragg angle increases from \ang{39.56} to \ang{54.44}. This leads to increased dispersion, and thus the angular encoder error effect is reduced from \SI{0.0036}{\electronvolt} in Ar to \SI{0.00168}{\electronvolt} in S. For the same reason, the effect of the temperature stabilisation contribution to the uncertainty is decreased. 

However, other contributions due to crystal structure are increased
due to the higher sensitivity to the shape of the simulated spectra. The contributions associated with the crystal structure and geometric effects are evaluated through simulations, following the procedure described in Ref.~\cite{assg2014}. All geometric effects (crystal tilts, vertical divergence and variation of x-ray source size) yield a higher energy shift in S than in Ar. This means that for higher dispersion there is an increase in sensitivity to possible misalignments of the instrument and to the unknown spatial location of the ions in the plasma. Thus, these effects have an higher impact on the final uncertainty for larger Bragg angles. Due to this increase in sensitivity to the spectrum shape,  the uncertainties related to crystal structure are also larger for S than for Ar. The simulated spectra considering different Si crystal form factor databases are more disparate at this lower energy, thus yielding higher deviations in energy depending on which one is used. We thus use the difference in the results obtained with the three different form factor databases as the uncertainty for this effect. A detailed investigation into the effect of form factors is presented in Section~\ref{sec:form_facts}. A similar procedure is followed for the uncertainty coming from a possible linear polarization of x rays. This effect is estimated by considering two different sets of simulations, one with an unpolarized reflectivity profile and another obtained with a $\sigma$ polarized reflectivity profile. The difference in transition energy obtained via the analysis using the different sets of simulations is considered as the uncertainty due to the presence of any polarized x-rays.

We have also added another source of uncertainty in the analysis of the energy of this line. As mentioned above, the Doppler broadening is evaluated as a weighted average value of the one obtained for each individual dispersive spectrum. Therefore, we checked for a possible energy dependence in the analysis due to the Doppler width value used in the simulations. For the evaluation of this effect, we performed the data analysis with simulations with the average value for the Doppler width and with simulations with the average value $+\sigma$ and $-\sigma$. The largest energy deviation to the analysis considering the average value of the Doppler broadening has been added to the uncertainty budget.
Because of the contribution of these effects, the final relative uncertainty is increased from \SI{2.5}{ppm} in Ar to \SI{4}{ppm} in S.

\begin{table}[tb]
\textcolor{black}{
	\caption{\label{tab:syst_uncert}
	Contributions to the systematic uncertainties for the M1 transition in sulfur. All energies are given in \si{\electronvolt}.
	}
	\begin{ruledtabular}
		\begin{tabular}{ld}
			\multicolumn{1}{c}{Contribution}  & \multicolumn{1}{c}{Value} \\  
			\hline
			Angular encoder error (\ang{;;0.2})  & 0.00171  \\
			Temperature stabilization (\SI{0.5}{\celsius})   & 0.00312  \\
			Vert. tilts of crystals (\ang{+-0.01}) for each crystal & 0.00085\\
			Vert. divergence (\SI{1}{\milli\meter}) & 0.00102\\
   			Variation of x-ray source size \SIrange{6}{12}{\milli\meter} & 0.00462\\
			Si crystal atomic form factor & 0.00300 \\
   			X-ray polarization   & 0.00513  \\
   			Lattice spacing error   & 0.00010  \\
   			Index of refraction   & 0.00055  \\
      		Thermal expansion   & 0.00015  \\
        	Energy dependence of the Doppler width   & 0.00031 \\
            \\
			Total  & 0.00819 \\
		\end{tabular}
	\end{ruledtabular}
	}
\end{table}

\section{Comparison with Theory}
\label{sec:comp_theo}
\begin{figure}[tb]
\includegraphics[width=\linewidth]{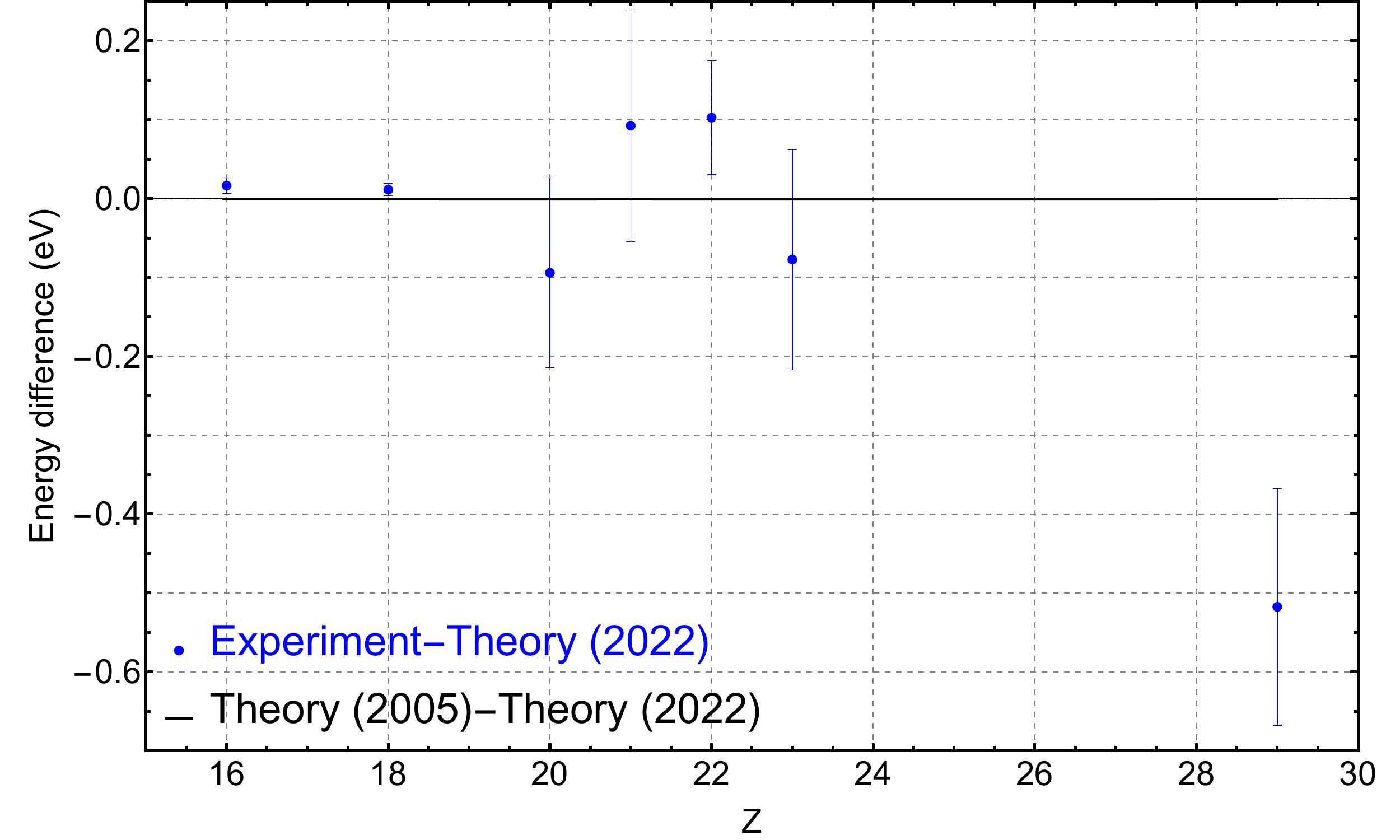}
\caption{Comparison between experiment and theory for the $1s 2s \; ^{3}S_{1}\to$ . $Z=16$: this work, $Z=18$ \cite{asgl2012}, $Z=20$: \cite{rrag2014}, $Z=21$: \cite{rgtm1995}, $Z=22$: \cite{pckg2014}, $Z=23$: \cite{cphs2000}, $Z=29$: \cite{bab2015}. Theory (2005): \cite{asyp2005}, Theory (2022): \cite{ypp2022}.}
\label{fig:comp-theory-exp}
\end{figure}

Our experimental transition energy as  compared with most advanced BSQED calculations are shown in Table~\ref{tab:breakdown}. This is the first ever measurement of this transition in sulfur, and also the most precise measurement of a $ n=2 \to n=1$ transition in this element. Our result is particularly interesting to compare with theory, as this intermediate $Z$ region is at the interface where both perturbative methods with respect to the $Z \alpha$ parameter, best adapted for light-$Z$ ions, and non-perturbative methods, best adapted for high-$Z$ species, may both be used, but each is at the limits of its domain of applicability. It is also in this region where unaccounted for contributions from each of these methods may reach their maximum, see discussion in \cite{yap2010}, thus precision experiments able to test these methods are essential. The theoretical results from Artemyev  \etal~\cite{asyp2005} include the complete set of two-electron QED corrections of order $\alpha$ and $\alpha^2$ evaluated to all orders in $Z$, a method well adapted to high-$Z$ species and whose accuracy is tested here for this medium-$Z$ ion. The very recent calculations from Yerokhin \etal~\cite{ypp2022}, building on their previous results \cite{yap2010} is based on the unified approach of Drake \cite{dra1988}, which aims at bridging the gap between perturbative and non-perturbative calculations for medium-$Z$ species, now with improvements to account for higher-order QED effects of order $m \alpha^{7+}$. Both calculations yield transition energies for this $1s 2s \; ^{3}S_{1} \to 1s^{2} \; ^{1}S_{0}$ line that are lower than our experimental result, $1.8\sigma$ and $1.7\sigma$ for \cite{asyp2005} and \cite{ypp2022}, respectively. This same trend was observed when our group measured this same transition in He-like Argon \cite{asgl2012}. The comparison between existing experiment for the M1 transitions for $16\leq Z \leq 29$ and recent calculations are presented in Fig.~\ref{fig:comp-theory-exp}. The comparison of the theoretical values from Artemyev \etal ~\cite{asyp2005} and \cite{ypp2022} is presented on the same figure.

\begin{table}[tb]
	\caption{\label{tab:breakdown}
	Comparison between experimental transition energy and theoretical values (\si{\electronvolt}). Calculated contributions to the $1s^{2}\; ^{1}S_{0}$ and $1s 2s \; ^{3}S_{1}$ levels are from Ref. \cite{asyp2005}.}
	\begin{ruledtabular}
		\begin{tabular}{ldddc}
Contribution  	&	 \multicolumn{1}{c}{$1s^{2}\; ^{1}S_{0}$ }	&	\multicolumn{1}{c}{ $1s 2s \; ^{3}S_{1}$} 	&	 \multicolumn{1}{c}{Transition} 	&		\\
\hline									
$\Delta E_{\mathrm{Dirac}}$ 	&	-3495.0044	&	-874.5000	&	2620.5044	&		\\
$\Delta E_{\mathrm{int}}$ 	&	270.4822	&	80.9665	&	-189.5157	&		\\
$\Delta E^{\textrm{QED}}_{1\;\textrm{el}}$ 	&	0.7562	&	0.1014	&	-0.6548	&		\\
$\Delta E^{\textrm{QED}}_{2\;\textrm{el}}$ 	&	-0.0715	&	-0.0110	&	0.0605	&		\\
$\Delta E^{\textrm{QED}}_{\textrm{h.o.}}$ 	&	0.0009	&	0.0002	&	-0.0007	&		\\
$\Delta E_{\textrm{rec}}$ 	&	0.0563	&	0.0137	&	-0.0426	&		\\
Theo. \cite{asyp2005} 	&	-3223.7803	&	-793.4292	&	2430.3511	&	(3)	\\
Theo. \cite{ypp2022}	&		&		&	2430.35208	&	(89)	\\
Exp. (this work) 	&	 	&	 	&	2430.3685	&	(97)	\\
\end{tabular}
	\end{ruledtabular}
\end{table}

\section{Sensitivity to atomic form factors}
\label{sec:form_facts}
The M1 lines studied in this work have negligible natural line width, thus they are a good probe to explore the systematic effects in our experiment linked to the experimental response function. Together with the Gaussian Doppler broadening as described in Sec. \ref{sec:doppler}, the remaining line shape results from the incident x rays reflecting off the Si(111) crystals within our experimental geometry. For a photon scattering elastically on an atom (in this case the Si atoms), the scattering process can be broken up into three different processes: nuclear Thomson and nuclear resonance scattering, bound electron scattering also known as Rayleigh scattering, and Delbruck scattering, which accounts for vacuum fluctuations in the Coulomb field of the nucleus. As nuclear and vacuum fluctuation effects only become significant at high energies and large angles, the contributions that play a role in our case are photoabsorption and forward-angle Rayleigh scattering \cite{kzrg1995,cha1995}. 

For ease of use, resonant scattering amplitudes are generally described using atomic Form Factors (FF) \cite{hgd1993}. The FF is the Fourier transform of the electron distribution of the atom, and is generally written as the sum of three terms: an angular dependent term $f_{0}$, and two angular independent terms $f_{1}$ and $f_{2}$ which account for energy-dependent reflection and absorption, respectively. $f_{1}$ and $f_{2}$ are known as anomalous scattering factors, and may also be expressed as $f'$ and $f''$ depending on the notation, which can be related to $f_{1}$ and $f_{2}$ via a constant fractor. The available FF databases are based on data, $S$-matrix theory, or some combination of the two. Average discrepancies between different theories are ~$10-30\%$, but may be much larger near absorption edges. Below a few \si{\kilo\electronvolt}, in the region of interest here, the available data to constraint FF from photoabsorption measurements have large experimental uncertainties, and are thus unable to discriminate between different FF models \cite{cha2000}, hence the interest of testing the FF sensitivity in this measurement. In a more recent work, the mass absorption coefficient of silicon has been measured with improved accuracy \cite{tcb2003} and the difference between this measurements and tabulated and theoretical values are of the order of a few percent.

\begin{figure}[tb]
\includegraphics[width=\linewidth]{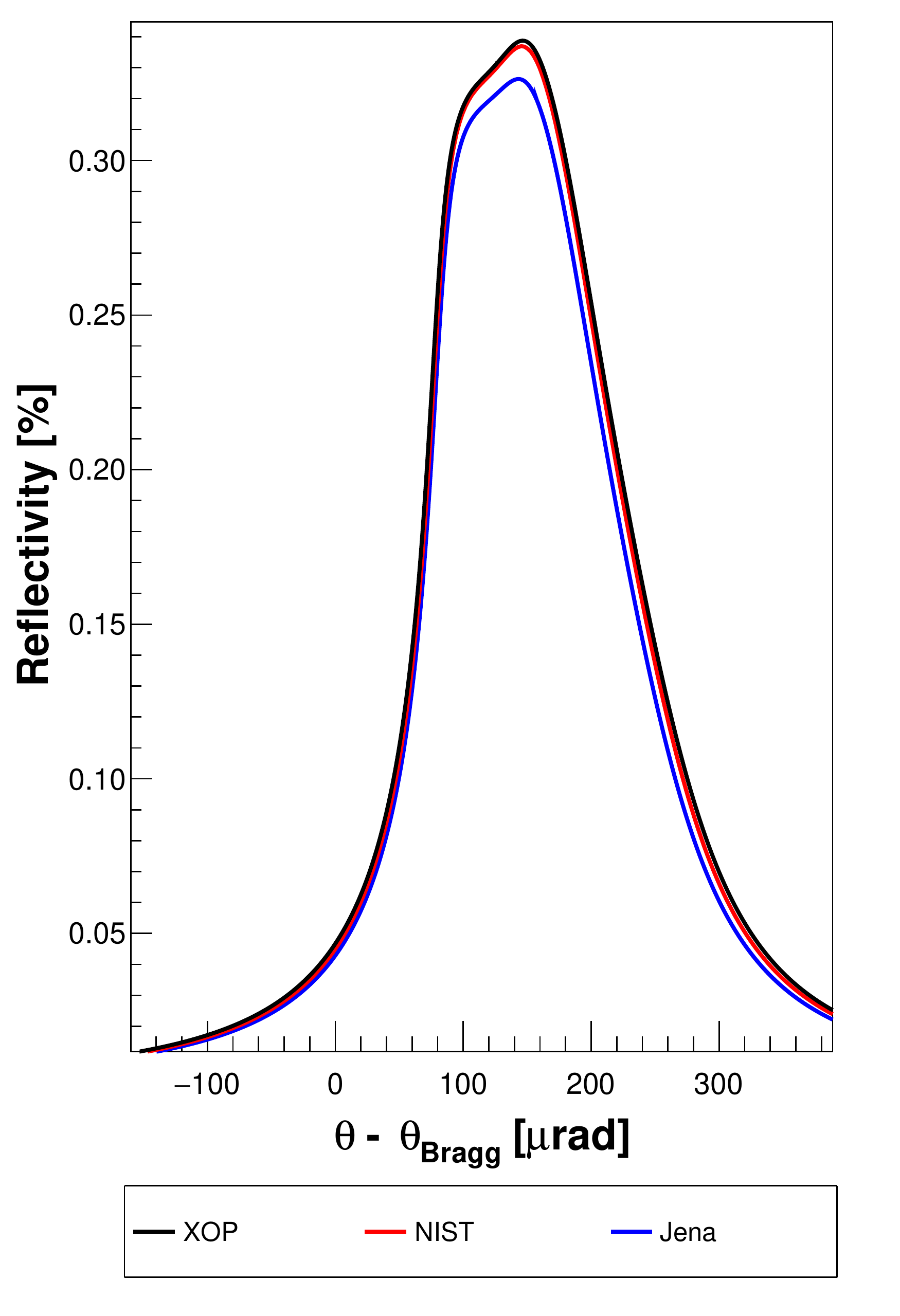}
\caption{ Si(111) Rocking curves for unpolarized radiation $\frac{1}{2}(\pi \; \mathrm{polarized} + \sigma  \; \mathrm{polarized})$ calculated with the different models used to produce the DCS response functions. The curves have been evaluated for the theoretical energy of the  $1s 2s \; ^{3}S_{1} \to 1s^{2} \; ^{1}S_{0}$ transition of 2430.3511 eV taken from Ref.~\cite{asyp2005}.}
\label{fig:Rocking_curves}
\end{figure}

Sensitivity to the FF was tested by generating experimental response functions using rocking curves obtained from the Jena, XOP-Henke, and NIST codes. NIST refers to the McXtrace program \cite{bpbt2013} that uses the RTAB form factor database \cite{kis2000} from $S$-matrix calculations for the $f1$ and $f2$ components, and the NIST FFAST database for f0 \cite{cha1995, cha2000}. XOP uses the Henke database \cite{hgd1993}. The Jena model, DIXI \cite{hwf1998}, uses the Henke experimental database and theoretical values from Sasaki \cite{sas1989}. Si(111) rocking curves from these three models for the x-ray energy regime of interest are shown in Fig.~\ref{fig:Rocking_curves}. As explained previously, these rocking curves are used as inputs for our \textit{ab initio} simulation of our spectrometer used to generate the simulated instrumental response functions.  

\begin{figure}[tb]
\includegraphics[width=\linewidth]{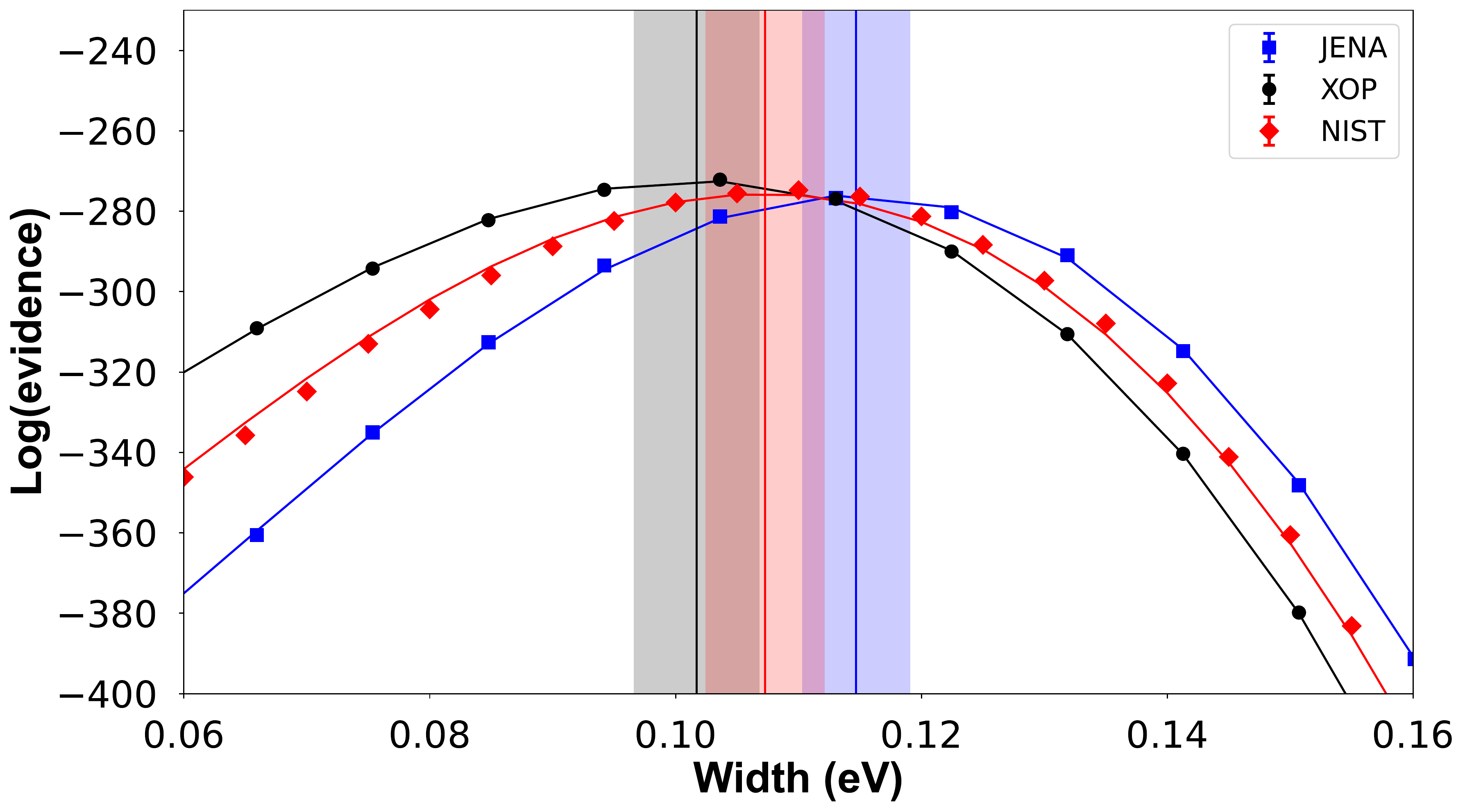}
\caption{Bayesian evidence (BE) curves for different Doppler widths for a single experimental dispersive spectra. Fits are shown with response functions generated with different atomic form factors models. The BE uncertainties are smaller than the size of the points. The BE trends as a function of width are shown fit with eight degree polynomials, from with the maximum evidence and standard BE uncertainty are obtained, shown by the solid lines and colored bars, respectively.}
\label{fig:S_M1_FF_dopp}
\end{figure}

The data analysis was then performed as described in Secs.~\ref{sec:doppler} and~\ref{sec:transE}. Doppler broadenings were simulated in the range \SIrange{0}{116}{\milli\electronvolt} FWHM. The experimental dispersive spectra were fitted with these simulated spectra using the Bayesian analysis toolkit NestedFit \cite{tra2019}. For each experimental spectrum and FF model, the Bayesian evidence was obtained as a function of the Doppler width, the results fit with an eighth-order polynomial and the evidence maximum and change in log(evidence) of \num{0.96} used to determine the Doppler broadening and associated uncertainty for each spectrum. The results for a single dispersive spectrum are shown in Fig.~\ref{fig:S_M1_FF_dopp}, where it is clear that the maximum shifts slightly for the different FF models. XOP and NIST models yield Doppler widths consistent within the uncertainties, but the results from the Jena model are not compatible with XOP, and the Jena model yields a larger Doppler width. This may be understood by examining the line shapes in Fig.~\ref{fig:Rocking_curves}, where it is clear that the rocking curve obtained with the Jena model is narrower than the others.

\begin{table}[tb]
\textcolor{black}{
	\caption{\label{tab:FF_comp}
	Comparison of Doppler widths and transition energies determined using different atomic form factor models. All energies are given in \si{\electronvolt}.
	}
	\begin{ruledtabular}
		\begin{tabular}{lcr}
			Model & Observable & Value  \\  \hline
			& Doppler width & \\
			XOP  & & 0.093(7)  \\
			NIST & & 0.100(7)  \\
			JENA & & 0.108(6)  \\ \hline
			& Energy &  \\ 
			XOP  & & 2430.3685(97)  \\
			NIST  & & 2430.3672(97)  \\
			JENA  & &  2430.3654(97)  \\
		\end{tabular}
	\end{ruledtabular}
	}
\end{table}

\begin{figure}[tb]
\includegraphics[width=\linewidth]{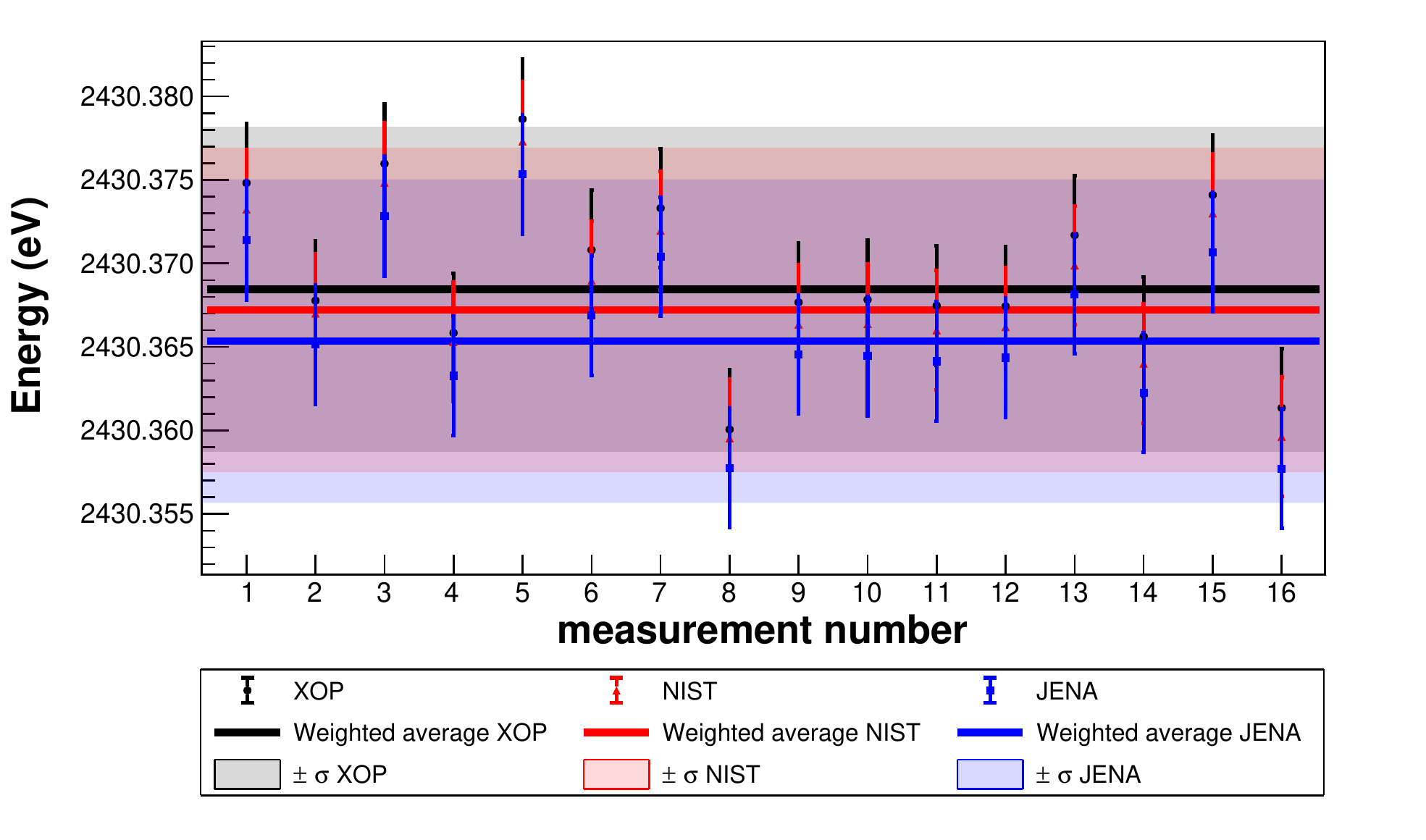}
\caption{ Comparison of the transition energy results between analysis using different models for the reflective profile calculation. The error bars on the points correspond to the statistical, temperature and the angle encoder measure uncertainties added in quadrature. The black dots correspond to the energy values extracted in the analysis using simulations performed with XOP~\cite{sad1998,sad2004,sad2011} calculated profiles, the red triangles using NIST profiles~\cite{bpbt2013} and the blue squares using JENA~\cite{hwf1998} profiles. The black, red and blue solid lines represent the weighted average of the different energy values using the respective points of same color. In the weighted average calculation of the energy for each model, only the statistical uncertainty of each point is considered. The shaded areas corresponds to the statistical uncertainty.}
\label{fig:S_M1_En_Diff_Models}
\end{figure}

Using these Doppler widths, the energy analysis was then performed following the procedure in Section~\ref{sec:transE}. The results with the averages over the different pairs of experimental spectra are shown in Fig.~\ref{fig:S_M1_En_Diff_Models}. Unlike the evaluation of the Doppler widths, the average transition energies extracted with the different FF models are compatible within the statistical uncertainties. This shows that these FF cannot be a significant source of uncertainty in the comparison with the BSQED calculations, confirming the systematic uncertainty related to the FF included in Table~\ref{tab:syst_uncert}. 

We note that there is a statistically significant difference in absolute value of Bayesian evidence for the different FF models, \ie it is possible to determine which line shape corresponds most closely with our data, but a full study of this effect is beyond the scope of this analysis. A summary of the Doppler widths and transition energies obtained with the different FF models is shown in Table~\ref{tab:FF_comp}.

\section{Conclusion}

We have made the first measurement of the $1s 2s \; ^{3}S_{1} \to 1s^{2} \; ^{1}S_{0}$ transition in He-like sulfur. The measurement was performed using the Paris double crystal spectrometer without any external energy reference. The measured transition energy is \SI{2430.3685+-0.0097}{\electronvolt}, which is in agreement with recent most advanced QED calculations within $2\sigma$. With an accuracy of \SI{4}{ppm}, this is the second more accurate measurement of the He-like M1 transition energy after the one on Ar with a relative accuracy of \SI{2.5}{ppm}.  The extremely narrow natural linewidth of the transition allowed us to perform a complementary study of the effect of the choice of the Si crystal atomic form factor model in our experimental response functions, as we use them here for the analysis in a region that is poorly constrained experimentally. We found that although the different models yield different Doppler broadenings, this does not have a significant impact in our transition energy analysis. Further work will aim at extending our analysis of He-like ions to higher-$Z$ species, in complement to ongoing studies of He-like ions at GSI.
\vspace{0.2cm}

%======= ACKNOWLEDGMENTS =============================================
%
\begin{acknowledgments}
%\section{Acknowledgements}
 We acknowledge  support from the PESSOA Huber Curien Program 2022, Number 47863UE and the PAUILF program 2017-C08. 
This research was funded in part by by Fundação para a Ciência e Tecnologia (FCT) (Portugal) through the research center grant UID/FIS/04559/2020 to LIBPhys-UNL from the FCT/MCTES/PIDDAC, Portugal.
Laboratoire Kastler Brossel (LKB) is ``Unit\'e Mixte de Recherche de Sorbonne Universit\'e, de ENS-PSL Research University, du Collège de France et du CNRS n$^{\circ}$ 8552''. 
P.I. and N.P. are members of the Allianz Program of the Helmholtz Association, contract n$^{\circ}$ EMMI HA-216 ``Extremes of Density and Temperature: Cosmic Matter in the Laboratory''.
The SIMPA ECRIS has been financed by grants from CNRS, MESR,  and  University Pierre and Marie Curie (now Sorbonne Universit\'e). The DCS has been constructed using  grants from BNM \textit{01 3 0002} and the ANR \textit{ANR-06-BLAN-0223}.
N.P. thanks the CNRS for support. L.D. thanks the Sorbonne University Institute \textit{Physics of Infinities} for his PhD grant. 
We wish to thank Dr. Marcus Mendenhall for valuable discussions regarding the NIST code.
%\vspace{3cm}
\end{acknowledgments}

\bibliographystyle{apsrev}

\bibliography{refs2022.bib}

\end{document}